\documentclass[runningheads]{llncs}

\usepackage{amssymb}
\usepackage{graphicx}

\usepackage{url}
\newcommand{\keywords}[1]{\par\addvspace\baselineskip
\noindent\keywordname\enspace\ignorespaces#1}

\newcommand\CHS{{\scshape Chisio}}
\newcommand\ChiBE{{\scshape ChiBE}}

\begin{document}

\mainmatter

\title{Chisio: A Compound Graph\\
Editing and Layout Framework}

\titlerunning{Chisio}

\author{Cihan Kucukkececi\and Ugur Dogrusoz\thanks{To whom correspondence 
should be addressed (\email{ugur@cs.bilkent.edu.tr})}\and Esat Belviranli\and
Alptug Dilek}
\authorrunning{Kucukkececi et al}

% the affiliations are given next
\institute{i-Vis Research Group, Computer Eng. Dept.,\\
Bilkent University, Ankara 06800, Turkey\\
%\mail\\
\url{http://www.cs.bilkent.edu.tr/~ivis/chisio}}

\maketitle

\begin{abstract}
  We introduce a new free, open-source compound graph editing and layout framework named \CHS, based on the Eclipse Graph Editing Framework (GEF) and written in Java. \CHS\ can be used as a finished graph editor with its easy-to-use graphical interface. The framework has an architecture suitable for easy customization of the tool for end-user's specific needs as well. \CHS\ comes with a variety of graph layout algorithms, most supporting compound structures and non-uniform node dimensions. Furthermore, new algorithms are straightforward to add, making \CHS\ an ideal test environment for layout algorithm developers. 


\end{abstract}

%%%%%%%%%%%%%%%%%%%%%%%%%%%%%%%%%%%%%%%%%%%%%%%%%%%%%%%%%%%%%%%%%%%%%%%%%%%%%%%%%%
\section{Introduction}
  As graphical user interfaces have improved, and more state-of-the-art software
  tools have incorporated visual functions, interactive graph editing and 
  layout facilities have become important components of software systems. Over 
  the years, an abundant number of such tools have been made available for 
  consumption both commercially and academically~\cite{KW01,DETT99,DFMDF02}.
  However, only a
  few, if any, non-commercial systems seem to address compound structures of 
  graphs both in regards to editing and layout capabilities. \CHS\ should fill 
  an important gap in this field.

  \CHS\ can be used as a finished graph editor and layout tool. The tool has a
  user-friendly graphical interface for interactive editing of compound graphs,
  accepting graphs in GraphML format~\cite{GraphML}. It also features a number 
  of graph layout algorithms from spring embedders to Sugiyama algorithm
  for hierarchical graphs. In addition, many aspects of the tool, from graph 
  object UIs to menus to persistency operations may be easily customized for 
  specific applications.
  
	There has been a great deal of work done on general graph 
	layout~\cite{KW01,DETT99} but considerably less on layout of compound graphs, 
	which is a notion that has been in use to represent more complex types of 
	relations or varying levels of abstractions in	data~\cite{SM95}. The limited
	work done on compound graph layout has mostly focused on layout of hierarchical
	graphs~\cite{SM91,SG96,EFL96}, where underlying relational information is 
	assumed to be under a certain hierarchy. \CHS\ provides an implementation of a
	relatively recent spring embedder based layout algorithm for undirected 
	compound graphs, with arbitrarily deep nesting relations~\cite{CoSE09}. In 
	addition, some other well-known layout methods have been extended to support
	compound structures in \CHS.
	
	Finally, \CHS\ hosts an implementation of a new spring embedder based 
	circular layout algorithm for clustered graphs.
	
%%%%%%%%%%%%%%%%%%%%%%%%%%%%%%%%%%%%%%%%%%%%%%%%%%%%%%%%%%%%%%%%%%%%%%%%%%%%%%%%%%
\section{Graph and Drawing Model of \CHS}
\begin{figure}[htb]
  \centering
  \includegraphics[width=1.48in]{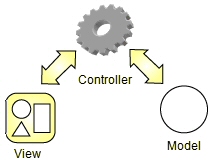}
  \caption{Model-view-controller structure used in GEF~\cite{GEF}}
  \label{fig:GEF}
  \end{figure}

	\CHS\ was built on top of the Eclipse Graph Editing Framework (GEF)~\cite{GEF}.
	GEF assumes that you have a model that you would like to display and edit
	graphically. The controllers bridge the view and model (Figure~\ref{fig:GEF}). 
	Each controller is responsible both for mapping the model to its view, and for
	making changes to the model. The controller also observes the model, and updates
	the view to reflect changes in the model's state. Controllers are the objects 
	with which the user interacts. 

\begin{figure}[htb]
  \centering
  \includegraphics[width=\textwidth]{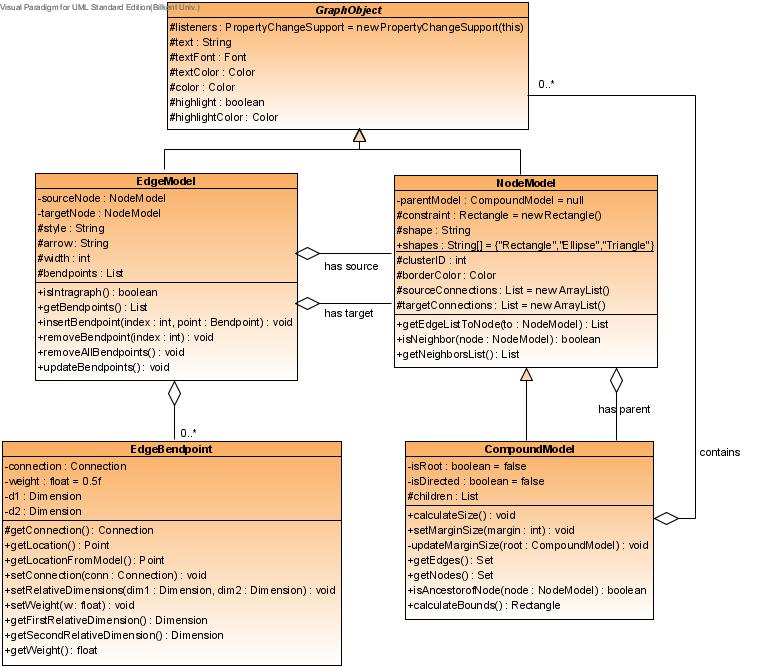}
  \caption{A UML class diagram illustrating the compound graph implementation of
  	\CHS}
  \label{fig:Chisio-mdl}
  \end{figure}

	The compound graphs are modeled in \CHS\ using this framework and the concept
	of graph managers described earlier (Figure~\ref{fig:Chisio-mdl}). A compound 
	node manages a list of children graph objects. A dedicated one is set as the 
	root of the nesting hierarchy. Through recursive use of compound nodes as child
	objects, an arbitrary level of nesting can be created.

\begin{figure}[htb]
  \centering
  \includegraphics[width=4in]{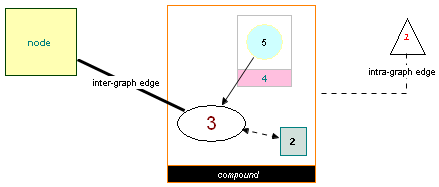}
  \caption{Basics of drawings in \CHS}
  \label{fig:Chisio-draw}
  \end{figure}

	Nodes, edges and compound nodes in \CHS\ all have distinct properties and UIs,
	which can be changed by its graphical user interface or through programming.
	Each node is drawn as a rectangle, ellipse or triangle. Edges can be drawn in
	a variety of styles. Compound nodes are always drawn with a rectangle, where
	the name text is displayed on its bottom margin and its geometry is 
	auto-calculated using the geometry of its contents to tightly bound its 
	contents plus user-defined child graph margins. Figure~\ref{fig:Chisio-draw} 
	shows the basics of drawing compound graphs in \CHS.

%%%%%%%%%%%%%%%%%%%%%%%%%%%%%%%%%%%%%%%%%%%%%%%%%%%%%%%%%%%%%%%%%%%%%%%%%%%%%%%%%%
\section{Chisio as an Editor}
	\CHS\ can be used as a finished graph editing and layout 
	tool~\cite{ChisioGuides}. Graphs created with other tools (and saved in GraphML
	format~\cite{GraphML}) and loaded into \CHS\ or graphs created in \CHS\ can be 
	edited and laid out interactively. 
	
\begin{figure}[htb]
  \centering
  \includegraphics[width=\textwidth]{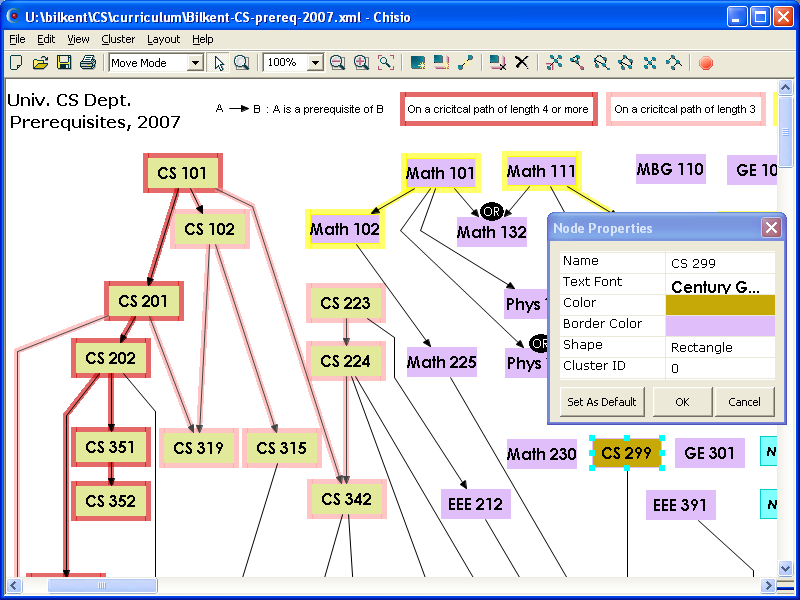}
  \caption{Example drawing in \CHS\ for course prerequisites in a department}
  \label{fig:Chisio-ss2}
  \end{figure}

\begin{figure}[htb]
  \centering
  \includegraphics[width=\textwidth]{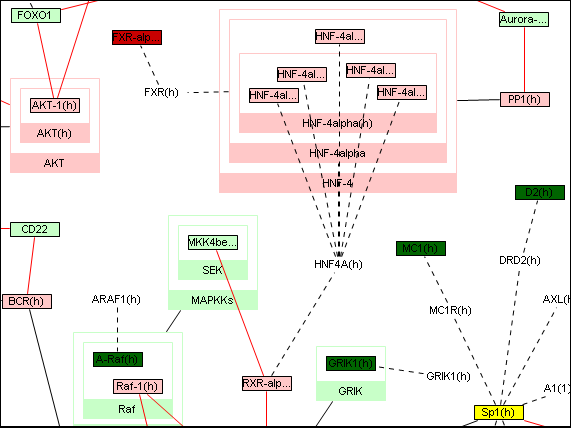}
  \caption{Partial example drawing in \CHS\ for a biological pathway}
  \label{fig:Chisio-ss3}
  \end{figure}

	The tool features standard graph editing facilities such as zoom, scroll, 
	add or remove graph objects, move, and resize. Object property and layout 
	options dialogs are provided to modify existing graph object properties and 
	layout options, respectively. In addition, printing or saving the current
	drawing as a static image and persistent storage facilities are supported.
	Furthermore, a highlight mechanism is provided to emphasize subgraphs of 
	user's interest. Figure~\ref{fig:Chisio-ss2} shows a sample screen-shot of \CHS.
	Drawing in~\ref{fig:Chisio-ss3} is a partial drawing produced by \CHS\ for 
	a biological pathway.

%%%%%%%%%%%%%%%%%%%%%%%%%%%%%%%%%%%%%%%%%%%%%%%%%%%%%%%%%%%%%%%%%%%%%%%%%%%%%%%%%%
\section{Customizing Chisio}
	One can customize \CHS\ for their specific needs by adding new node/edge types
	or by modifying existing nodes/edges (UI and attributes)~\cite{ChisioGuides}. 
	In addition, you may customize the menus to add new functionality as well as 
	modifying node and edge menus and property inspectors. Furthermore, \CHS\ is 
	designed for easy integration of new layout algorithms. Layout researchers will
	especially find \CHS\ to be useful for implementation and testing of their new
	methods.
	
\begin{figure}[htb]
  \centering
  \includegraphics[width=\textwidth]{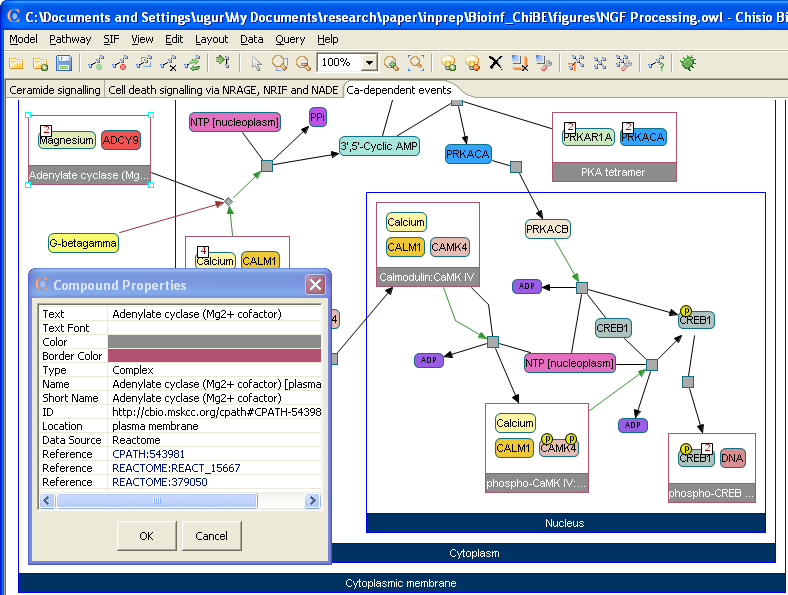}
  \caption{Sample screen-shot from \ChiBE}
  \label{fig:ChiBE-ss}
  \end{figure}

	A sample application based on Chisio is \ChiBE\ (Chisio BioPAX Editor) (\url{http://sourceforge.net/projects/chibe/}).
	The tool features user-friendly display, viewing, and editing of pathway models
	represented by the BioPAX format (\url{http://biopax.org}).
	Pathway views are rendered in a feature-rich format, and may be laid out
	automatically (Figure~\ref{fig:ChiBE-ss}). Among other things, visualization 
	of experimental data overlaid on pathway views is supported in \ChiBE.
	
%%%%%%%%%%%%%%%%%%%%%%%%%%%%%%%%%%%%%%%%
\subsection{Layout Architecture in Chisio}
\begin{figure}[htb]
  \centering
  \includegraphics[width=8cm]{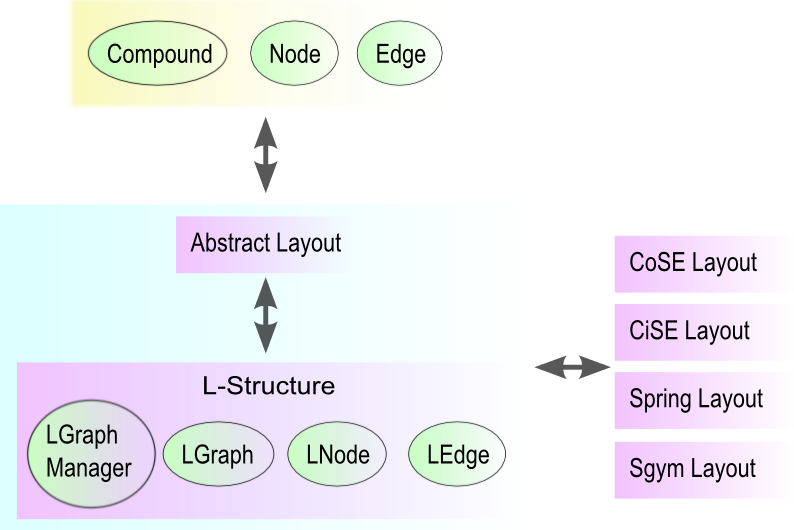}
  \caption{Layout architecture of \CHS}
  \label{fig:Chisio-layout}
  \end{figure}

	The basis in \CHS\ layout is constructed through an abstract layout class 
	named \verb+AbstractLayout+ and an associated l-structure,
	classes \verb+LGraphManager+, \verb+LGraph+, \verb+LNode+, and \verb+LEdge+.
	Here an \verb+LGraphManager+ maintains and manages a list of \verb+LGraph+
	instances, which in turn maintains a list of \verb+LNode+'s and \verb+LEdge+'s. 
	The \verb+AbstractLayout+ class converts the given \CHS\ model into a generic
	l-structure used only for layout purposes that is destroyed at the end of 
	layout. Individual layout algorithms extend the \verb+AbstractLayout+ class 
	and run on this l-structure by overriding predefined methods. When the layout 
	is finished, the geometry information of the l-level is transferred to \CHS\ 
	model (Figure~\ref{fig:Chisio-layout}).

%%%%%%%%%%%%%%%%%%%%%%%%%%%%%%%%%%%%%%%%%%%%%%%%%%%%%%%%%%%%%%%%%%%%%%%%%%%%%%%%%%
\section{Layout Algorithms in Chisio}
	\CHS\ provides implementations of the following layout methods.
	
%%%%%%%%%%%%%%%%%%%%%%%%%%%%%%%%%%%%%%%%
\subsection{CoSE Layout}
	CoSE (Compound graph Spring Embedder) layout is an algorithm specifically
	designed for compound graphs~\cite{CoSE09}. It has been based on the 
	traditional force-directed layout scheme with extensions to handle multi-level
	nesting, varying node sizes, and possibly other application-specific 
	constraints.
	
\begin{figure}[b]
  \centering
  \includegraphics[width=6cm]{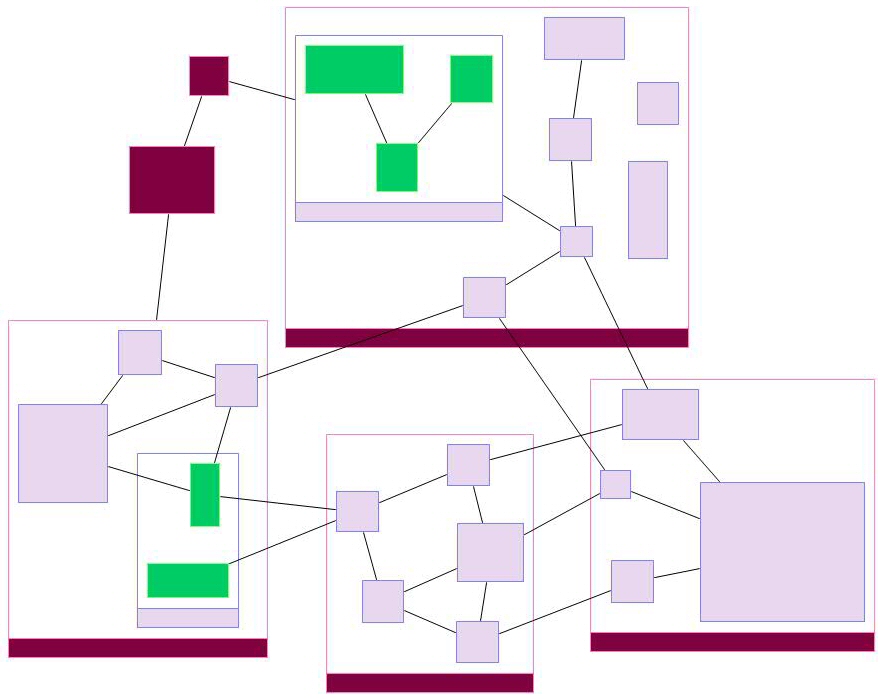}
  \caption{Sample drawing produced by the CoSE algorithm}
  \label{fig:CoSE}
  \end{figure}

	An expanded node and its associated nested graph are represented as a single
	entity, similar to a ``cart'', which can move freely in orthogonal directions
	(no rotations allowed). Multiple levels of nesting is modeled with smaller 
	carts on top of larger ones. Figure~\ref{fig:CoSE} shows an example drawing
	produced by this layout algorithm.

%%%%%%%%%%%%%%%%%%%%%%%%%%%%%%%%%%%%%%%%
\subsection{Cluster Layout}
\begin{figure}[htb]
  \centering
  \includegraphics[width=6cm]{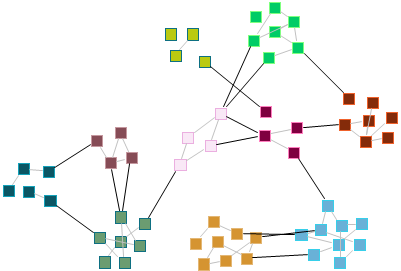}
  \caption{Sample drawing produced by the Cluster layout algorithm}
  \label{fig:Cluster}
  \end{figure}

	Cluster Layout is designed to place the nodes that belong to the same
	group or cluster (say to refer to molecules with similar functionality or 
	network devices in a LAN) near each other without the use of compound nodes
	(Figure~\ref{fig:Cluster}). It uses CoSE layout as a subroutine to do that.

%%%%%%%%%%%%%%%%%%%%%%%%%%%%%%%%%%%%%%%%
\subsection{CiSE Layout}
	A popular way to draw clustered graphs is in a circular fashion. In other
	words, a circle of appropriate size is created for each cluster and the 
	nodes in that cluster are placed around this circle trying to minimize 
	edge crossings. Circular layout algorithms~\cite{ME88,ST99,HS04,BB04} 
	address the issue of placing nodes of a cluster nicely around a circle, 
	minimizing the number of edge crossing and total edge lengths. However most
	of these algorithms do not address the issue of how the cluster graph should
	be laid out (i.e. how the individual circles should be placed with respect
	to each other to minimize crossings and lengths of inter-cluster edges).
	The ones that do, can not handle non-tree structures nicely. CiSE layout 
	algorithm does not place any constraints on the structure of the cluster
	graph and tries to satisfy the following specific requirements:
	\begin{itemize}
	\item Nodes in the same cluster should be placed around the same circle
	spaced as evenly as possible.
	\item Unclustered nodes should not be placed on or overlapping any circles,
	but they might have neighbor nodes placed around a circle.
	\item The number of (intra-cluster) edge crossings between the nodes in the
	same cluster should be as small as possible.
	\item The number of (inter-cluster) edge crossings between the nodes in
	different clusters or unclustered nodes should be as small as possible.
	\end{itemize}
  \begin{figure}[htb]
  \centering
  \includegraphics[width=8.4cm]{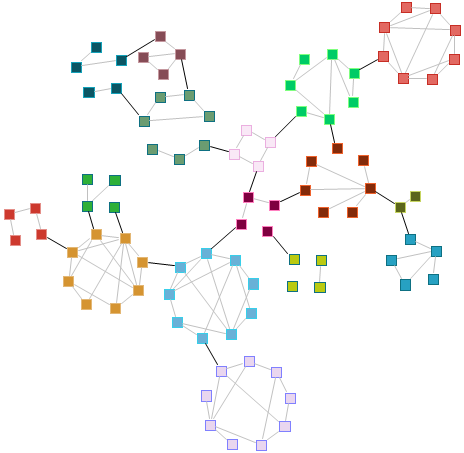}
  \caption{Sample drawing produced by the CiSE layout algorithm; nodes on
  each cluster and unclustered nodes are distinctly colored.}
  \label{fig:CiSE}
  \end{figure}
	
	The layout algorithm is based on the spring embedder, extending it 
	for properly grouping nodes in the same cluster and placing them around a 
	circle by extra constraints. The use of extra constraints is implemented by
	introducing additional properties to the physical model used by the spring 
	embedder, trying to obey the basic laws of physics.
	
	Each cluster/circle is represented by a meta-node of circular shape, on 
	which sits a round-shaped track on the periphery. The physical entities for
	each member node of a cluster is assumed to be either fixed (restrained to
	its current location on the owner circle) or flexible to move around the 
	track on which they sit as needed by the different steps of the algorithm. 
	In addition, we assume a dynamic center of gravity in the middle of the 
	bounding rectangle of the current drawing, towards which an attractive force
	is assumed by unclustered nodes and cluster nodes (all nodes except member 
	nodes of a cluster).

	Following are the main steps of the algorithm:
	\begin{itemize}
	\item {\bf Step 1}: In this step, we run a basic circular layout~\cite{HS04}
	for each cluster. 
	\item {\bf Step 2}: Next we layout the cluster graph, where each node in the
	graph represent a cluster with the dimensions produced in Step 1, using a
	basic spring embedder. 
	\item {\bf Step 3}: Here, our aim is to reposition/rotate circles according
	to the location of their out-nodes and inter-cluster edges incident on 
	these nodes. However, nodes on the circles are not allowed to move 
	individually. They are assumed to be pinned down to their circles. After 
	this step, a draft layout of the whole graph is obtained. 
	\item {\bf Step 4}: In this final polishing step, we obtain the final layout 
	of the graph by trying to reduce inter-cluster edge crossing. The main
	difference between this step and Step 3 is that we allow nodes on circles to 
	move on their parent circle (as well as moving with them).
	\end{itemize}

	\CHS\ features a draft-implementation of this algorithm and
	Figure~\ref{fig:CiSE} shows a sample drawing produced by this algorithm.

%%%%%%%%%%%%%%%%%%%%%%%%%%%%%%%%%%%%%%%%
\subsection{Others}
	\CHS\ also provides implementations for a regular spring embedder~\cite{KK}, 
	a hierarchical layout method~\cite{STT}, and a circular layout
	algorithm~\cite{HS04}.
	
%%%%%%%%%%%%%%%%%%%%%%%%%%%%%%%%%%%%%%%%%%%%%%%%%%%%%%%%%%%%%%%%%%%%%%%%%%%%%%%%%%
\section{Conclusion}
	In this paper we have presented a new open-source, free compound graph editing 
	and layout tool, distributed under Eclipse Public License. \CHS\ version 1 is
	available at \url{http://www.cs.bilkent.edu.tr/~ivis/chisio.html}.

\bibliography{reference}

\begin{thebibliography}{10}

\bibitem{BB04}
M.~Baur and U.~Brandes.
\newblock Crossing reduction in circular layouts.
\newblock In {\em 30th Intl. Workshop Graph-Theoretic Concepts in
  Computer-Science (WG '04)}, LNCS 3353, pages 332--343, Berlin, 2004.

\bibitem{GraphML}
U.~Brandes, M.~Eiglsperger, I.~Herman, M.~Himsolt, and M.~S. Marshall.
\newblock {GraphML} progress report.
\newblock In {\em Graph Drawing}, pages 501--512, 2001.

\bibitem{ChisioGuides}
{C}hisio {U}ser's and {P}rogrammer's {G}uides.
\newblock \url{http://www.cs.bilkent.edu.tr/~ivis/chisio.html}.
\newblock {i-Vis} Research Group, Bilkent University, Ankara, Turkey.

\bibitem{DETT99}
G.~{Di Battista}, P.~Eades, R.~Tamassia, and I.~G. Tollis.
\newblock {\em Graph Drawing, Algorithms for the Visualization of Graphs}.
\newblock Prentice-Hall, 1999.

\bibitem{DFMDF02}
U.~Dogrusoz, Q.~Feng, B.~Madden, M.~Doorley, and A.~Frick.
\newblock Graph visualization toolkits.
\newblock {\em IEEE Computer Graphics and Applications}, 22(1):30--37,
  January/February 2002.

\bibitem{CoSE09}
U.~Dogrusoz, E.~Giral, A.~Cetintas, A.~Civril, and E.~Demir.
\newblock A layout algorithm for undirected compound graphs.
\newblock {\em Information Sciences}, 179:980--994, 2009.

\bibitem{EFL96}
P.~Eades, Q.~Feng, and X.~Lin.
\newblock Straight-line drawing algorithms for hierarchical graphs and
  clustered graphs.
\newblock In S.~North, editor, {\em Graph Drawing (Proc. GD '96)}, volume 1190
  of {\em Lecture Notes in Computer Science}, pages 113--128. Springer-Verlag,
  1997.

\bibitem{GEF}
{G}raphical {E}diting {F}ramework version 3.1.
\newblock The Eclipse Foundation, 2007.
\newblock http://www.eclipse.org/gef.

\bibitem{HS04}
H.~He and O.~Sykora.
\newblock New circular drawing algorithms.
\newblock In {\em Workshop on Information Technologies - Applications and
  Theory (ITAT'04)}, 2004.

\bibitem{SM95}
{K. Sugiyama and K. Misue}.
\newblock {A Generic Compound Graph Visualizer/Manipulator: D-{ABDUCTOR}}.
\newblock In F.~J. Brandenburg, editor, {\em Graph Drawing (Proc. GD '95)},
  volume 1027 of {\em Lecture Notes in Computer Science}, pages 500--503.
  Springer-Verlag, 1995.

\bibitem{KK}
T.~Kamada and S.~Kawai.
\newblock An algorithm for drawing general undirected graphs.
\newblock {\em Information Processing Letters}, 31(1):7--15, April 1989.

\bibitem{KW01}
M.~Kaufmann and D.~Wagner, editors.
\newblock {\em Drawing Graphs: Methods and Models}, volume 2025 of {\em Lecture
  Notes in Computer Science}. Springer, 2001.

\bibitem{ME88}
E.~Makinen.
\newblock On circular layouts.
\newblock {\em International Journal of Computer Mathematics}, 24:24--29, 1988.

\bibitem{SG96}
G.~Sander.
\newblock Layout of compound directed graphs.
\newblock Technical Report A/03/96, University of Saarlandes, CS Dept.,
  Saarbr{\"{u}}cken, Germany, 1996.

\bibitem{ST99}
J.~M. Six and I.~G. Tollis.
\newblock Circular drawings of biconnected graphs.
\newblock In {\em 1st Workshop Algorithm Engineering and Experimentation
  (ALENEX '99)}, volume 1619 of {\em Lecture Notes in Computer Science}, pages
  57--73. Springer-Verlag, 1999.

\bibitem{SM91}
K.~Sugiyama and K.~Misue.
\newblock Visualization of structural information: Automatic drawing of
  compound digraphs.
\newblock {\em IEEE Transactions on Systems, Man and Cybernetics},
  21(4):876--892, 1991.

\bibitem{STT}
K.~Sugiyama, S.~Tagawa, and M.~Toda.
\newblock Methods for visual understanding of hierarchical systems.
\newblock {\em IEEE Transactions on Systems, Man, and Cybernetics},
  21(2):109--125, February 1981.

\end{thebibliography}
\bibliographystyle{abbrv}

\end{document}